\documentclass{article}
\usepackage{latexsym,ifsym,tocbibind}
\usepackage[utf8]{inputenc} 
\usepackage[T1]{fontenc}
\usepackage{url}
\usepackage{cite}
\usepackage{deleq,color}
\usepackage{relsize, exscale}
\usepackage{amsmath,amssymb,amsfonts,amscd,graphicx}

\usepackage{psfrag}
\usepackage{epstopdf}

{\catcode `\@=11 \global\let\AddToReset=\@addtoreset}
\AddToReset{equation}{section}
\renewcommand{\theequation}{\thesection.\arabic{equation}}

{\catcode `\@=11 \global\let\AddToReset=\@addtoreset}
\AddToReset{figure}{section}

{\catcode `\@=11 \global\let\AddToReset=\@addtoreset}
\AddToReset{table}{section}

\newcommand{\secS}{\blue{S}}
\newcommand{\blue}[1]{{\color{blue}#1}}

\newcommand{\ph}{\Phi}
\newcommand{\ps}{\Psi}
\newcommand{\upsi}{\underline{\Psi}{}}

\newcommand{\novarphibutpsi}{\red{\psi}}
\newcommand{\nopsibutvarphi}{\red{\varphi}}

\newcommand{\ptcheck}[1]{\ptc{checked on #1}}

\newcommand{\mypi}{{\color{blue}{\pi}}}

\newcommand{\red}[1]{{\color{red} #1}}

\newcommand{\rmnote}[1]{}

\newcommand \al {\alpha}
\newcommand \bet {\beta}

\newcommand{\bean}{\begin{eqnarray}\nonumber}
\newcommand{\beal}[1]{\begin{eqnarray}\label{#1}}
\newcommand{\eeal}[1]{\label{#1}\end{eqnarray}}
\newcommand{\bel}[1]{\begin{equation}\label{#1}}

\newcommand{\hJ}{{\hat J}}

\newcommand{\ts}{{\cal S}} 

\newcommand{\hxi}{\hat \xi}
\newcommand{\hxib}{\hat{\xib}}
\newcommand{\heta}{\hat \eta}
\newcommand{\hetab}{\hat{\etab}}
\newcommand{\home}{\hat{\omega}}
\newcommand{\homb}{\hat{\underline\omega}}

\newcommand{\oube}{\,\,\mathring{\!\!\ubeta}}
\newcommand{\oubeta}{\oube}

\newcommand{\hups}{\hat{\ups}}
\newcommand{\hupsb}{\hat{\upsb}}

{\catcode `\@=11 \global\let\AddToReset=\@addtoreset}

\AddToReset{figure}{chapter}

\newcommand{\tr}{\mbox{tr}}

\newcommand{\be}{\begin{equation}}
\newcommand{\bea}{\begin{eqnarray}}
\newcommand{\eea}{\end{eqnarray}}
\newcommand{\beaa}{\begin{eqnarray*}}
\newcommand{\eeaa}{\end{eqnarray*}}
\newcommand{\bseq}{\begin{subeq}}
\newcommand{\eseq}{\end{subeq}}
\newcommand{\eql}[1]{\arrlabel{#1}}
\newcommand{\alp}{\alpha}

\def \rectangle#1#2{\hbox{\vrule\vbox to #2
{\hrule\hbox to
#1{\hfil}\vfil\hrule}\vrule}}
\newcommand{\edd}{\end{document}}
\newcommand{\lot}{\mbox{l.o.t.}}
\newcommand{\dd}{\mbox{${\hD}$}}

\newcommand{\nabb}{\mbox{$\nabla \mkern-13mu /$\,}}
\newcommand{\hnabb}{\mbox{$\hnabla \mkern-13mu /$\,}}

\newcommand{\nn}{\nonumber}
\newcommand{\nnn}{}
\newcommand{\wdd}{d} 

\newcommand{\chib}{\underline{\chi}{}}
\newcommand{\hchi}{\hat{{\chi}}}
\newcommand{\hchib}{\hat{\underline{\chi}}{}}
\newcommand{\Psib}{\underline{\Psi}{}}

\newcommand{\zetab}{\underline{\zeta}{}}
\newcommand{\hzeb}{\hat{\zetab}}

\newcommand{\half}{\frac{1}{2}} 

\newcommand{\xib}{\underline{\xi}}

\newcommand{\und}[1]{\underline{#1}}

\newcommand{\vb}{\und{v}}

\newcommand{\M}{{\mycal M}}

\newcommand{\ua}{{\underline{\alpha}}{}}
\newcommand{\ualpha}{{\ua}}
\newcommand{\ualp}{{\ua}}
\newcommand{\bb}{{\underline{\beta}}{}}
\newcommand{\ub}{{\bb}{}}
\newcommand{\ubeta}{{\bb}{}}

\newcommand{\si}{\sigma}

\newcommand{\ro}{\rho}

\newcommand{\ze}{\zeta}
\newcommand{\hze}{\hat\zeta}
\newcommand{\divv}{\mbox{div}\mkern-19mu /\,\,\,\,}
\newcommand{\hdivv}{\widehat{\mathrm{div}}\mkern-19mu /\,\,\,\,}

\newcommand{\om}{\omega}
\newcommand{\oom}{\omega}

\newcommand{\omb}{\underline{\oom}}

\newcommand{\etab}{\underline{\eta}}

\newcommand{\ddd}{\nabb}

\newcommand{\hdddd}{{\bf \hat D} \mkern-13mu /\,}

\newcommand{\uchi}{\underline{\chi}}

\def\frac#1#2{{{#1}\over{#2}}}

\newcommand{\ab}{\underline{\alpha}}

\newcommand{\eeq}{\end{equation}}
\newcommand{\ee}{\end{equation}}
\newcommand{\beqa}{\begin{eqnarray}}
\newcommand{\beqar}{\begin{deqarr}}
\newcommand{\beqarn}{\begin{deqarr}\nonumber}
\newcommand{\beqarl}[1]{\begin{deqarr}\label{#1}}
\newcommand{\eeqa}{\end{eqnarray}}
\newcommand{\eeqar}{\end{deqarr}}
\newcommand{\eeqarl}[2]{\label{#1}\arrlabel{#2}\end{deqarr}}
\newcommand{\beqan}{\begin{eqnarray*}}
\newcommand{\eeqan}{\end{eqnarray*}}
\newcommand{\ba}{\begin{array}}
\newcommand{\ea}{\end{array}}

\newtheorem{theorem}{Theorem}[section]

\DeclareFontFamily{OT1}{rsfs}{} \DeclareFontShape{OT1}{rsfs}{m}{n}{
 <-7> rsfs5 <7-10> rsfs7 <10-> rsfs10}{}
\DeclareMathAlphabet{\mycal}{OT1}{rsfs}{m}{n}

\newcounter{mnotecount}[section]

\renewcommand{\themnotecount}{\thesection.\arabic{mnotecount}}

\newcommand{\mnote}[1]
{\protect{\stepcounter{mnotecount}}$^{\mbox{\footnotesize
$
\bullet$\themnotecount}}$ \marginpar{
\raggedright\tiny\em $\!\!\!\!\!\!\,\bullet$\themnotecount: #1} }

\newcommand{\ptc}[1]{\mnote{{\bf PTC}: #1}}

\newcommand{\R}{\mathbb R}

\newcommand{\eq}[1]{(\ref{#1})}

\newcommand{\Eq}[1]{Equation~\eq{#1}}

\newcommand{\matr}{\mathring}

\newcommand{\tcp}[1]{ { \color{magenta} #1} }

\renewcommand{\ptcheck}[1]{}
\renewcommand{\red}[1]{#1}
\renewcommand{\checkmark}{}
\renewcommand{\tcp}[1]{#1}

\begin{document}

\title{The ``neighborhood theorem'' for the general relativistic characteristic Cauchy problem in higher dimensions\thanks{Preprint UWThPh-2023-15}}

\author{Piotr T. Chru\'sciel\thanks{University of Vienna,
Faculty of Physics, Boltzmanngasse 5, A 1090 Vienna}\\
Roger Tagne Wafo\thanks{
D\'epartement de Math\'ematiques et Informatique, Facult\'e des
Sciences, Universit\'e de Douala}
\\
 Finnian Gray$^\dagger$}

\maketitle

\tableofcontents
\begin{abstract}
We show that the maximal globally hyperbolic solution of the initial-value problem for the  higher-dimensional
vacuum Einstein equations on two transversally intersecting characteristic
hypersurfaces contains a future neighborhood of the hypersurfaces.
\end{abstract}

\section{Introduction}

A classical theorem of Rendall~\cite{RendallCIVP2} shows that characteristic initial data for the vacuum Einstein equations on two three-dimensional characteristic hypersurfaces intersecting transversally at a submanifold $S$ lead to a vacuum spacetime that covers a future neighborhood of $S$. The result has been extended to higher dimensions in~\cite{CCM2}.
It has been shown in~\cite{Luk} that the maximal developement of the data covers in fact a whole neighborhood of the initial data hypersurfaces when $S$ is a sphere, and the same result has been established in~\cite{CCW} for all topologies of $S$. The object of this work is to generalise this result to all higher dimensions.

For this we check that the vacuum Einstein equations, possibly with a cosmological constant, can be written in a form to which the results in~\cite{CCW} apply. We note that there are at least two avenues of approach
here. The first is to use the usual wave-coordinates reduction of the Einstein equations. There arises, however,  an apparent difficulty, that the estimates developed in~\cite{CCW}
are based on a doubly-null coordinate system, but the transition to such coordinates from harmonic ones can lead to a loss of two derivatives. While this does not allow one to use directly the estimates of~\cite{CCW},  a tame version of these together with the Nash-Moser iteration scheme can be used to solve the problem. We will not provide any details of such an argument here but  present instead an approach based on Bianchi identities, which has the potential to be useful for other
future problems. The idea goes back to an observation of
Senovilla~\cite{SenovillaSymHyp}, that the Bianchi identities also lead to a symmetric hyperbolic system of equations in higher dimensions. It then remains to check that, similarly to the four-dimensional problem, these equations can be written in a form to which suitable estimates apply.
In effect, we show that the Christodoulou-Klainerman version~\cite{ChristodoulouKlainerman93} of the four-dimensional Newman-Penrose formalism has an obvious counterpart in higher dimensions.%
\footnote{Compare~\cite{ReallHighDimNP,PravdaHighDimNP} and references therein.}
This allows us to prove:

\begin{theorem}
  \label{T7IV23.1}
The maximal globally hyperbolic development  of smooth initial data for  vacuum Einstein equations with a (possibly zero) cosmological constant $\Lambda \in \R$,  on two characteristic hypersurfaces intersecting transversally, contains a future neighborhood of the initial data hypersurfaces.
\end{theorem}

We have formulated the theorem in the smooth category for simplicity, but the result remains true in suitable Sobolev spaces, as can be checked by chasing various losses of differentiability in the reformulation of Einstein equations as a system to which the estimates of \cite{CCW} apply.

A formal proof of Theorem~\ref{T7IV23.1} can be found in Section~\ref{s8IV23.1}.
To prove this theorem we  need the principal part of a subset of the Bianchi equations, presented in Appendix~\ref{App26IV23.1}. For future reference we give the full  form of all Bianchi equations in Appendix~\ref{App28IV23.1}.

{\color{red}
Once this paper was written we found that the system of equations considered here has already been derived in~\cite{RodnianskiShlapentokh} for  related  purposes (compare~\cite{Collingbourne}). An alternative proof of our Theorem~\ref{T7IV23.1} is given as Theorem~A.1 there with a minor difference, namely compactness of the intersection is assumed. In our approach this hypothesis is not necessary, using standard domain-of-dependence arguments.
}

\section{Doubly-null systems}
 \label{s5IV23.1}

 We will write the vacuum Einstein equations in a form suitable for the estimates of \cite{CCW}, namely
 \begin{equation}\label{5IV23.1}
   A^\mu \partial_\mu f  = \lot
   \,,
 \end{equation}
 with $(x^\mu=(u,v,x^A)$, $A=2,\ldots,n$,
where $\lot$ denotes a smooth function of $f$ and of the coordinates, and where the coefficients $A^\mu$ are allowed to depend upon $f$. For this the field $f$ should be written in the form
\bel{bid2.5}f= \left(%
 \begin{array}{c}
  \varphi \\
  \psi \\
 \end{array}%
\right)%
\,, \ee
with
\bel{bid3} A^u=\left(%
\begin{array}{cc}
  A^u_{\varphi\varphi} & 0 \\
  0 & 0 \\
\end{array}%
\right) \ \mbox{and}\
 A^v=\left(%
\begin{array}{cc}
  0 & 0 \\
  0 & A^v_{\psi\psi} \\
\end{array}%
\right) \
 \mbox{
satisfying $A^v_{\psi\psi} > 0$, $A^u_{\varphi\varphi}> 0$.}
\ee
Thus the only non-vanishing components of $A^u$ are contained in the block
$A^u_{\varphi\varphi}$, and those of $A^v$ are contained in the block
$A^v_{\psi\psi}$.

The key assumption in \cite{CCW} was the existence of a scalar product $\langle \cdot,\cdot\rangle$ so that
\begin{equation}\label{5IV23.3}
  \partial_\mu \big( \langle f_1, A^\mu f_2\rangle \big)= \lot
\end{equation}
for all fields $f_1$ and $f_2$.
This will be the case when the matrices $A^\mu$ are symmetric with respect to $\langle \cdot,\cdot\rangle$, as assumed in \cite{CCW}. The main point of this work is to establish this.

\subsection{Doubly null frames}
\label{sSWcdnf}

We start by  describing our framework. We follow closely the notation and presentation in \cite{CCW},
which in turn follows~\cite{KlainermanNicoloBook}.

Consider any field of vectors $e_{i}$  such that
\be
(g_{ij}):=(g(e_i,e_j))=\left(\begin{array}{ccc}  0& -2 & 0 \\
-2& 0 & 0 \\ 0 & 0 & \delta^a_b
 \\ \end{array} \right)\,,
 \label{metric}
\ee
where indices $i,j$ {\em etc.\/} run from $0$ to $n$, while
indices $a,b$ {\em etc.\/} run from $2$ to $n$. One therefore has
$$ (g^{ij}):=
g(\theta^i,\theta^j)=\left(\begin{array}{ccc} 0 & -1/2 & 0 \\
-1/2 &0 & 0
\\  0& 0 & \delta^a_b  \end{array} \right)\,,
$$
where $\{\theta^{i}\}$ is a basis of $T^*\M$ dual to $\{e_{i}\}$. If
$u_i$ is the usual Lorentzian orthonormal basis
of $T\M$,
$$g(u_i, u_j)=\eta_{ij}=\mathrm{diag}(-1, 1,\ldots, 1)\,,
$$
then a basis $\{e_i\}$ as above can be constructed by setting
$$
e_a=u_a\,, \quad e_0=u_0+u_1\,, \quad e_1=-u_0+u_1\,.
$$

We let
$$
 \ts=\mathrm{Vect}(\{e_2, \ldots, e_n\})\,,
$$
where Vect$(\Omega )$ denotes the vector space spanned by the elements of
the set $\Omega$.

\newcommand{\hDlocal  }{D}%
\newcommand{\hGlocal }{\Gamma}%
For any connection $\hDlocal  $ we define the connection
coefficients $\hGlocal _\red{i}{}^\red{j}{}_\red{k}$ by the formula
$$
\hGlocal _\red{i}{}^\red{j}{}_\red{k}:=\theta^\red{j}(\hDlocal _{e_\red{i}}e_\red{k})\,,$$
so that
$$
 \hDlocal _{e_\red{i}}e_\red{k} = \hGlocal _\red{i}{}^\red{j}{}_\red{k} e_\red{j}
 \,.
$$
The connection $\hDlocal  $ has no torsion if and only if
$$
\hDlocal  _{e_\red{i}}e_\red{k} -\hDlocal  _{e_\red{k}}e_\red{i} =
[e_\red{i},e_\red{k}]\,,
$$
and it is  metric compatible if and only if
\begin{equation}\label{metnotcomp}
\hDlocal  _{\red{i}}g_{\red{j} \red{k}}\equiv (\hDlocal
_{e_\red{i}}g)(e_\red{j},e_\red{k}) = -\hGlocal _{\red{i}\red{j}\red{k}}-\hGlocal
_{\red{i}\red{k}\red{j}}=0\,.
\end{equation}
Here and elsewhere, $$\hGlocal _{\red{i}\red{j}\red{k}}:=g_{\red{j}\red{\ell} }\hGlocal
_\red{i}{}^\red{\ell}{}_\red{k}\,.$$


\subsection{The doubly-null decomposition of Weyl-type  tensors}
 \label{ss9V14.1}
%
%

Let $\wdd^i{}_{jkl}$  be any tensor field with the symmetries of the
Weyl tensor, \be\label{weylsyms} \wdd_{ijkl}= \wdd_{klij}\,, \quad
\wdd_{ijkl}= -\wdd_{jikl}\,, \quad g^{jk}\wdd_{ijkl}=0\,, \quad
\wdd_{i[jkl]}= 0\,.
\ee
We decompose $\wdd^i{}_{jkl}$ into its null
components, relative to the null pair $\{e_0,e_1\}$, as follows:%
\footnote{In spacetime dimension four, our $e_0$ corresponds to $e_4$ in \cite{CCW}, while $e_1$ here is denoted by $e_3$ in this last reference.}
\begin{deqarr}
&&\ua(X,Y)=\wdd(X,e_1,Y,e_1)\,,\quad \alp(X,Y)=\wdd(X,e_0,Y,e_0)\,,\nnn\\
&&\ub(X)=\frac{1}{2}\wdd(X,e_1,e_1,e_0)\,,\quad
\beta(X)=\frac{1}{2}\wdd(X,e_0,e_1,e_0)\,,\nnn
 \\
 &&
 \ro=\frac{1}{4}\wdd(e_1,e_0,e_1,e_0)\,,
 \quad \sigma(X,Y):= \frac 12 d(X,Y,e_1,e_0)
 \,,
 \nnn
 \\
 &&
 \ph(X,Y)= \wdd(X,e_1,Y,e_0)
 \,,
 \\
&& \ps(X,Y,Z)= d(X,Y,Z, e_0)\,, \nnn\quad \upsi(X,Y,Z)= d(X,Y,Z,e_1)
\,,
 \eql{decomp}
 \\
&& \pi(V,X,Y,Z)= d(V,X,Y,Z)\,,
 \eql{decompPi}
\end{deqarr}
where $V, X,Y$ and $Z$ are arbitrary vector fields orthogonal to $e_0$ and
$e_1$.
The fields $\alpha$ and $\ua$ are symmetric and traceless. From
\eq{decomp} together with the symmetries of the Weyl tensor one finds
\begin{deqarr}
\wdd_{a1b1}=\ua_{ab}\,, & \wdd_{a0b0}= \alp_{ab}\,, \nnn\\
\wdd_{a110}=2\ub_{a}\,, & \wdd_{a010}= 2\beta_{a}\,, \nnn\\
\wdd_{1010}=4\rho_{}\,, & \qquad \wdd_{ab10}=2\si_{ab}=  2\Phi_{[ab]}\,, \nnn\\
\wdd_{abc1}=\upsi_{abc}\,, & \wdd_{abc0}= \ps_{abc}
 \,, \nnn\\
\wdd_{a1b0}=\ph_{ab}  \,, & \wdd_{abcd}=  \mypi_{abcd}
 \,.
 \arrlabel{identi}
\end{deqarr}
%
%

Furthermore,
\begin{equation}\label{23III23.91}
\Phi_{[ab]} = \checkmark   \sigma_{ab}
   \,,
   \quad
    \Psi_{[abc]} = \underline \Psi_{[abc]} = 0
  \,,
\end{equation}
The vanishing of the traces of the Weyl tensor gives
%
\begin{eqnarray}
 &
   \tr_h \alpha =  \checkmark \tr_h \underline \alpha =0
   \,,
   \
   \rho = \checkmark -\frac{1}{2} \tr_h \Phi
   \,,
   \
   \underline \beta_c = \checkmark h^{ab}\underline \Psi_{bca}
   \,,
   \
    \beta_c = \checkmark - h^{ab}  \Psi_{bca}
   \,,
   &
    \nonumber
\\
&
   \pi^c{}_{acb} \equiv h^{cd}\pi_{acbd}  = \checkmark  \Phi_{(ab)}
   \,,
   \
   \pi^{ab}{}_{ab} \equiv h^{ac} h^{bd}\pi_{abcd}  = \checkmark - 2 \rho
   \,.
   &
   \label{26III23.1}
\end{eqnarray}
In particular we see that $\rho$ is determined by a double trace of $\pi_{abcd}$, and does therefore not need to be considered as a separate variable.

The following symmetry rules are useful in deriving an explicit form of the equations:
\begin{equation}
  e_0\leftrightarrow e_1
  \,,
  \
   \alpha \leftrightarrow \underline \alpha
  \,,
  \
   \Psi \leftrightarrow \underline \ps
  \,,
  \
   \rho \leftrightarrow \rho
  \,,
  \
   \beta \leftrightarrow - \underline \beta
   \,,
  \
   \ph_{ab} \leftrightarrow \ph_{ba}
  \,,
  \
   \sigma_{ab} \leftrightarrow - \sigma_{ab}
  \,,
   \label{8IV23.1}
\end{equation}

\subsection{The equations}
\label{ssdndBe}

Recall the second Bianchi identity for a Levi-Civita connection
$D$,
\bel{12XI13.10}
 D_i R_{j k \ell m} +
 D_j R_{ k i \ell m} +
 D_k R_{i j\ell m} =
 0
 \,.
\ee
Contracting $ i $ with $m $ one obtains
\bel{12XI13.11}
 D_i R_{j k \ell}{}^i +
 D_j R_{k\ell } -
 D_k R_{j\ell} =
 0
 \,.
\ee
Inserting into this equation  the expression for the Riemann tensor
in terms of the Weyl and Ricci tensors,
\bea
 R_{jk\ell}{}^{i}
  = W_{j k\ell}{}^{i}
 + 2\left(g_{\ell[j} L_{k]}{}^{i}
 - \delta^{i }{}_{[j}L_{k]\ell} \right)
 \label{conf6}
\;
\end{eqnarray}
%
%
%
 where
\begin{equation}
 L_{ij} :=  \frac{1}{n-1}R_{ij} - \frac{1}{2n(n-1)} R g_{ij}
 \,,
\end{equation}
we obtain
\bel{Weyleq} D_i W^i{}_{jk\ell} =g^{i m}\hDlocal_i W{}_{m j k \ell}
\equiv J_{j k \ell}\,,
\ee
where
\bel{12XI13.14}  J_{j k \ell}
 :=
 \frac{2(n-2)}{n-1}D_{[k}R_{\ell]j}
-
 \frac{n-2}{n(n-1)}g_{j[\ell}D_{k]}R
  \,.
\ee
Here, and elsewhere, square brackets around a set of $\ell$ indices
denote antisymmetrization  with a multiplicative factor $1/\ell!$.

From now on we assume that we have a solution of the Einstein equations, possibly with a cosmological constant $\Lambda \ne 0$, thus the Ricci tensor is proportional to the metric and
\begin{equation}\label{8IV23.1ag}
  J_{j k \ell} \equiv 0
  \,.
\end{equation}

\section{Closed subsystems with symmetric hyperbolic principal part}
 \label{s6IV23.1}

 The Bianchi identities imply a set of equations for the Weyl tensor, the principal part of which is derived in Appendix~\ref{App26IV23.1}. (The full set of equations, the details of which are not needed for the proof of Theorem~\ref{T7IV23.1}, are presented in Appendix~\ref{App28IV23.1}.)
  From these equations we can derive a closed system of equations to which the estimates of~\cite{CCW} apply. We list and analyse the relevant subsystems in what follows.

The symbol ``$\lot$'' in the equations below denotes polynomials in the rotation coefficients and the tetrad components of the Weyl
 tensor, with coefficients which are rational functions of the tetrad.

\subsection{$\big(\ua,(\ub,\upsi)\big)$}
 \label{ss6IV23.1}

The coupled system for $ \ua\,, \;\ub\;$ and $\upsi$   reads
(Equations \eqref{04IV23.81a}, \eqref{23III23.81ba} and
\eqref{23III23.81h})
\begin{deqarr}
 e_0(\ua_{ab})&= \checkmark  & -
 \red{
        e_{(a} (\underline \beta_{b)})
        -e_c(\upsi^c{}_{(ab)})
         }
  + \lot,
   \\
 e_1(\underline \beta_a) &=\checkmark & - e_b(\ua_{a}{}^b)+  \lot,
  \\
e_1 (\underline \Psi_{abc}) &=\checkmark& -e_a ( \underline
\alpha{}_{bc})
     + e_b(\underline \alpha{}_{ac})
     +\lot
    \,.
     \arrlabel{6IV23.11}
\end{deqarr}

Let
\begin{equation}\label{6IV23.1}
 f= \left(
          \begin{array}{c}
            \nopsibutvarphi \\
            \novarphibutpsi \\
          \end{array}
        \right)
        ,
        \quad
  \novarphibutpsi = (\ua_{ab})\,,
  \quad
  \nopsibutvarphi= \left(
          \begin{array}{c}
            \ub_a \\
            \upsi_{abc} \\
          \end{array}
        \right)
         \,.
\end{equation}
We consider the scalar product
\begin{equation}\label{6IV23.2}
 \langle f, \hat f\rangle \equiv  \langle
   \left(
          \begin{array}{c}
            \nopsibutvarphi  \\
            \novarphibutpsi\\
          \end{array}
        \right)
        ,
        \left(
          \begin{array}{c}
            \hat\nopsibutvarphi  \\
            \hat\novarphibutpsi\\
          \end{array}
        \right)
         \rangle
        := \ua_{ab} \hat \ua{}^{ab} +
         \ub{}_{a} \hat \ub{}^{a} +
         \frac 12
         \upsi_{abc} \hat \upsi{}^{abc}
        \,.
\end{equation}
In order to check symmetric-hyperbolicity  we  rewrite \eqref{6IV23.11} as
\begin{deqarr}
 e_0(\ua_{ab})&=& -M^c{}_{ab} e_c(\nopsibutvarphi)
  + \lot
   ,
   \\
 e_1(\underline \beta{}_a) &= & - L^c{}_a e_c(\novarphibutpsi)
     +\lot,
  \\
e_1 (\underline \Psi_{abc}) &=& -L^d{}_{abc}e_d(\novarphibutpsi)
     +\lot
    \,.
     \arrlabel{6IV23.12}
\end{deqarr}
Since the left-hand side is diagonal, to check the symmetric-hyperbolic character of this system it suffices to check the symmetry of the symbol of the operator $ -A^ae_a$ appearing on the right-hand side. For $k\in T^*\secS$ we can write
\begin{eqnarray}
 A^\perp(k) := A^ak_a
 =
   \left(
    \begin{array}{cc}
      0 &  M(k)  \\
      L(k)  & 0 \\
    \end{array}
  \right)
  \,,
\end{eqnarray}
where the notation should be clear from \eqref{6IV23.12}. Then
\begin{eqnarray}
\lefteqn{
 \langle f,  A^\perp(k) \hat f\rangle
   \equiv
      \langle
   \left(
          \begin{array}{c}
            \nopsibutvarphi  \\
            \novarphibutpsi\\
          \end{array}
        \right)
        ,
        A^\perp(k)
        \left(
          \begin{array}{c}
            \hat\nopsibutvarphi  \\
            \hat\novarphibutpsi\\
          \end{array}
        \right)
         \rangle
         }
         &&
         \nonumber
\\
 &&
     =
         \ua{}^{ab} \big(k_a\hat \ubeta{}_b+k_c\hat\upsi{}^c{}_{ab}
         \big) +
         \ub{}^{a} k_b\hat\ua_{a}{}^b
          +
         \frac 12
         \upsi{}^{abc}\big( k_a   \underline {\hat\alpha}{}_{bc}
     - k_b \underline{\hat \alpha}{}_{ac}
      \big)
       \nonumber
\\
 &&
     =
         \ua{}^{ab}  k_a\hat \ubeta{}_b
        + \hat  \ua{}^{ab}  k_a   \ubeta{}_b
           +\ua{}^{ab}  k_c\hat\upsi{}^c{}_{ab}
           +\hat \ua{}^{ab}  k_c \upsi{}^c{}_{ab}
        \,.
        \label{6IV23.13}
\end{eqnarray}
The last line is obviously symmetric under interchange of $f$ and $\hat f$,
which shows that the system is indeed symmetric hyperbolic, as desired.

\subsection{$\big((\matr{\ub}\,, \matr{\upsi})\,, \ph\big)$}
 \label{ss6IV23.2}

In order to obtain a symmetric hyperbolic system for all components of the Weyl tensor it will be convenient to write a system of equations where some fields appear twice. For this we introduce
 $$
\matr\ub : =  \ub \qquad \mbox{and}\qquad \matr{\underline \Psi}=
 \underline \Psi \;.
 $$
 Should one need a direct existence theorem for the solution in the current setting, one would need to show that initial data for which $\matr\ub=\ub$ initially lead to solutions were this equality holds everywhere. However, we do not need an existence theorem, but a system of equations where suitable estimates apply. Indeed, we start with a solution of the vacuum Einstein equations, define all the field as above, and thus the equality $\matr\ub=\ub$ hold everywhere. In particular it does not matter in the ``$\lot$'' terms whether or not $\ub$ there is taken to be $\matr \ub$ or $\ub$. For definiteness we make the latter choice here, though other choices could be more convenient for other purposes.

%
 We put together Equations (\ref{23III23.81a}), (\ref{23III23.81i}
)  and (\ref{04IV23.81b}) and obtain the following closed system
\begin{deqarr}
e_0(\matr\ub_a)&= \checkmark&  e_b ( \Phi_a{}^b) +  \lot
  \,,
\\
 e_0(\matr{\underline \Psi}_{abc}) &= \checkmark&  e_b(\Phi_{ca} )  - e_a(\Phi_{cb}) +\lot
    \,,
    \\
 e_1(  \Phi_{ab} ) &= \checkmark&
        e_b (\underline {\matr\beta}_a)    -   e_c(\matr{ \underline \Psi}^c{}_{ba}) +\lot
     \,.
\end{deqarr}
Calculations identical to those in Section~\ref{ss6IV23.1} show that the principal part of this system is symmetric hyperbolic.

\subsection{$\big((\bet\,, \ps)\,, \al\big)$}
A closed system for $\bet\,, \ps $ and $ \al$ is given by
(\ref{23III23.81bac}),  (\ref{23III23.81h}) and (\ref{04IV23.81b}):
\begin{deqarr}
e_0(\bet_a)&= \checkmark & e_b(\al_a{} {}^b) +\lot
 \,,
\\
 e_0 (\ps_{abc}) &= \checkmark  & -e_a (   \alpha_{bc})
     + e_b(  \alpha_{ac})
     +\lot
    \,,
 \\
 e_1(  \alpha_{ab} ) & = \checkmark &
        e_{(a} (  \beta_{b)})    -   e_c \ps^c{}_{(ab)} +\lot
     \,.
\end{deqarr}
This system can be obtained by the symmetry operations \eqref{8IV23.1} from the system considered in Section~\ref{ss6IV23.1}, and symmetric-hyperbolicity easily follows.

\subsection{$(\sigma_{ab},\matr{\Psi}_{abc},\pi_{abcd})$}

We let
\begin{equation}\label{8IV23.11}
  \matr{\Psi}_{abc}:= \Psi_{abc}
\end{equation}
and consider the system of equations \eqref{24III23.14},
\eqref{23III23.81k3} and \eqref{23III23.71} for the fields
$(\sigma_{ab},\matr{\Psi}_{abc},\pi_{abcd})$:
\begin{deqarr}
    e_0(\sigma_{ab})
   &= \checkmark
   &
      e_c(  \matr{\Psi}_{ab}{}^c)
     +\lot
    \,,
   \label{P24III23.14a}
\\
  e_0(\pi_{abcd})  &= \checkmark&
  e_b (  \matr{\Psi}_{cda})
  -   e_a(  \matr{\Psi}_{cdb})
     +\lot
      \,,
   \label{P23III23.81k3a}
\\
  e_1(\matr{\Psi}_{abc})  &= \checkmark&
      e_c(\sigma_{ab})
  -    e_d(\pi^d{}_{cab})
     +\lot
   \label{P23III23.71a}
    \,,
    \arrlabel{6IV23.31x}
\end{deqarr}
as well as
\begin{equation}\label{P6IV23.1}
 f= \left(
          \begin{array}{c}
            \nopsibutvarphi \\
            \novarphibutpsi \\
          \end{array}
        \right)
        ,
        \quad
         \nopsibutvarphi = (\matr{\Psi}_{abc})\,,
  \quad
  \novarphibutpsi= \left(
          \begin{array}{c}
            \sigma_{ab} \\
            \pi_{abcd} \\
          \end{array}
        \right)
         .
\end{equation}
We consider the scalar product
\begin{equation}\label{P6IV23.2}
 \langle f, \hat f\rangle
        := \sigma_{ab} \hat \sigma{}^{ab} +
         \frac 12
         \pi_{abcd} \hat \pi{}^{abcd}+
         \matr{\Psi}{}_{abc} \hat {\matr{\Psi}}{}^{abc}
        \,.
\end{equation}
As before it suffices to check the symmetry of the symbol $A^\perp(k)$ of the operator $ -A^ae_a$ appearing on the right-hand side of \eqref{6IV23.31x}.  We find
\begin{eqnarray}
\lefteqn{
 \langle f,  A^\perp(k) \hat f\rangle
   \equiv
      \langle
   \left(
          \begin{array}{c}
            \nopsibutvarphi  \\
            \novarphibutpsi\\
          \end{array}
        \right)
        ,
        A^\perp(k)
        \left(
          \begin{array}{c}
            \hat\nopsibutvarphi  \\
            \hat\novarphibutpsi\\
          \end{array}
        \right)
         \rangle
         }
         &&
         \nonumber
\\
 &&
     =
         \sigma{}^{ab}  k_c\hat{\matr{\Psi}}{}_{ab}{}^c{}
          +
         \frac 12
         \pi{}^{abcd}\big( k_b   \hat{\matr{\Psi}}{}_{cda}
     - k_a \hat \matr{\Psi}{}_{cdb}
      \big)
           +
         \matr{\Psi}^{abc}
         (k_c\hat\sigma{}_{ab} - k_d \hat \pi{}^d{}_{cab}
         )
       \nonumber
\\
 &&
      =
         \sigma{}^{ab}  k_c\hat{\matr{\Psi}}{}_{ab}{}^c{}  +
         \matr{\Psi}^{abc}
          k_c\hat\sigma{}_{ab}
          +
         \pi{}^{abcd}  k_b   \hat{\matr{\Psi}}{}_{cda}
          +
         \matr{\Psi}^{cda}
          k^b \hat \pi{}_a{}^b{}_{cd}
        \,,
        \label{P6IV23.13}
\end{eqnarray}
which is symmetric under interchange of $f$ and $\hat f$,  as desired.

\section{The proof}
 \label{s8IV23.1}

We are ready to pass to the proof of Theorem~\ref{T7IV23.1}, which is essentially a repetition of the four-dimensional proof in \cite{CCW}.  We verify some  possibly-dimension-dependent steps of the argument for the convenience of the reader.

We will use a set of equations, emphasised by
Friedrich (see~\cite{FriedrichCargese} and references therein),
for a frame field $e_i= e_i{}^\mu \partial_\mu$:
\begin{deqarr}
&&[e_{p},e_{q}] = (\Gamma_{p}\,^{l}\,_{q} -
\Gamma_{q}\,^{l}\,_{p})\,e_{l}
 \,,\label{E1x}
\\
    &&e_{p}(\Gamma_{q}\,^{i}\,_{j}) - e_{q}(\Gamma_{p}\,^{i}\,_{j}) -
    2\,\Gamma_{k}\,^{i}\,_{j}\,\Gamma_{[p}\,^{k}\,_{q]}
         +
    2\,\Gamma_{[p}\,^{i}\,_{|k|} \Gamma_{q]}\,^{k}\,_{j}
    \nn
\\
 &&\quad =
    d^{i}\,_{jpq} +
    \frac{2}{n-1}
    \big(
      \delta^{ i}_{[p}R_{q]j} -  g_{j[p} R_{q]}{}^{i}
       \big)
 + \frac{2 R }{n(n-1)}  g_{j[p} \delta_{q]} ^{i}
 \label{conf6x}
 \,.
  \arrlabel{Hconf1Ex}
\end{deqarr}
Here the indices $i$, $j$, etc. are frame indices running from $0$ to $n$, with Greek indices being coordinate indices running over the same range.
The first equation \eq{E1x} says that $\Gamma$ has no torsion.
This, together with the requirement that the $\Gamma_{ijk}$'s be anti-symmetric
in the last two-indices  implies that
$\Gamma$ is the Levi-Civita connection of $g$. When
$d_{ijkl}$ has the symmetries of the Weyl tensor, the
left-hand-side of \eq{conf6x} is simply the definition of the
curvature tensor of the connection $\Gamma$, with $d_{ijkl}$ being
the Weyl tensor, $R_{ij}$ being the Ricci tensor and $R$ the Ricci
scalar.

Next, given a metric $g$, a standard construction (see, e.g., Section~4 of \cite{CCW}) gives a
coordinate system $(x^\mu)\equiv (u,v,x^A)$, $A\in\{2,\ldots,n\}$, where $\nabla u$ and $\nabla v$ are null, with the vector fields $e_i$,
$i=0,1,\ldots,n$, as in \eqref{metric} and of the form
\bel{14XI13.11}
 e_1 = \partial_u
 \,,
 \quad
 e_0 = e_0{}^v\partial_v + e_0{}^A\partial_A
 \,.
\ee
With this choice of tetrad,
\eq{E1x} becomes  an evolution equation for the tetrad coefficients
$e_i{}^\mu$
\bel{14XI13.15}
 [e_1,e_i] \equiv \partial_u e_i{}^\mu \partial_\mu = (\Gamma_1\,^{l}\,_{i} -
\Gamma_{i}\,^{l}\,_1)\,e_{l}
 \,.
\ee
By construction, the $\partial_u e_a$'s have no $u$ and $v$
components, which gives the identities
\bel{14XI13.21}
 0 = (\Gamma_1\,^{1}\,_{a} -
\Gamma_{a}\,^{1}\,_1)
 =(\Gamma_1\,^{0}\,_{a} -
\underbrace{\Gamma_{a}\,^{0}\,_1}_{=0})
 \,.
\ee
Next, $\partial_u e_0$ has no $\partial_u$-component, which implies
\bel{14XI13.25}
 0 = \underbrace{\Gamma_1\,^{1}\,_0}_{=0} -
 \Gamma_0\,^{1}\,_1
 \,.
\ee
Now, the vector fields $e_a$, $a=2,\ldots,n$ are determined up to
rotations in the space
 $\text{Vect}\{e_2,\ldots ,e_n\}$, and we can get rid
of part of this freedom by imposing
\bel{14XI13.12}
 \Gamma_1{}^a{}_b = 0
 \,.
\ee
Again by construction, the integral curves of the vector fields $e_1$ and
$e_0$ are null geodesics, though not necessarily
affinely-parameterised:
\bel{epc}  D_{e_1}e_1\sim e_1\,,
 \quad
D_{e_0}e_0\sim e_0\,. \ee
In this gauge
\bel{uxig}
 \Gamma_1{}^a{}_1 = 0=\Gamma_0{}^a{}_0
 \,.
\ee

The vanishing of the rotation coefficients just listed allows us to
get rid of the second term in some of the combinations
$$
 e_{1}(\Gamma_{q}\,^{i}\,_{j}) - e_{q}(\Gamma_1\,^{i}\,_{j})
$$
appearing in \eq{conf6x}. In this way, we can algebraically
determine
$$
 \partial_u \Gamma_{q}\,^{a}\,_1
 \
 \mbox{and}
 \
 \partial_u \Gamma_{q}\,^{a}\,_{b}
 %
$$
in terms of the remaining fields appearing in \eq{conf6x}. Similarly
we have
$$
 e_0(\Gamma_{q}\,^{a}\,_0) - e_{q}(\Gamma_0\,^{a}\,_0) =
 e_0(\Gamma_{q}\,^{a}\,_0)
 \,,
 \quad
 e_0(\Gamma_{q}\,^{1}\,_1) - e_{q}(\Gamma_0\,^{1}\,_1) =
 e_0(\Gamma_{q}\,^{1}\,_1)
 \,,
$$
which gives ODEs for $ e_0(\Gamma_{q}\,^{a}\,_0)$ and  $
e_0(\Gamma_{q}\,^{1}\,_1)$.

Keeping in mind \eq{14XI13.21} and the symmetries of the
$\Gamma_i{}^j{}_k$'s, we conclude that all the non-vanishing connection coefficients
satisfy ODEs along the integral curves of $e_1=\partial_u$ or of
$e_0$.

The analysis of the Bianchi equations in Section~\ref{s6IV23.1} leads in vacuum to  the following two
collections of fields,
\beal{phid}
\varphi&=&(e_i,\Gamma_i{}^a{}_b,\Gamma_i{}^a{}_1,\alp,\ubeta,
\underline\Psi,\matr{\Psi},\Phi)\,,\\
\psi&=&(\Gamma_i{}^a{}_0,\Gamma_i{}^3{}_1,\ualp,\oubeta,
\beta,\matr{\underline\Psi},\pi,\si,\Psi)
 \,,
\eeal{psid}
with the gauge conditions just listed,
\bean
 &
 e_1{}^u=1
 \,,
 \quad
 0=e_1{}^v=e_1{}^A=e_0{}^u=e_a{}^u=e_a{}^v
  \,,
  &
\\
& \Gamma_1{}^{1}{}_a= \Gamma_a{}^{1}{}_1
  \,,
  \quad
 0= \Gamma_1{}^a{}_1=  \Gamma_0{}^i{}_0
 = \Gamma_1{}^a{}_b
    \,,
   &
\eeal{15XI13.1}
to which the argument of the proof of Theorem~5.3 of \cite{CCW} applies.

\appendix

\section{Bianchi identities}
\label{App26IV23.1}

The algebraic and differential properties of the Riemann tensor of a vacuum metric lead to the following equations where, in case of more than one equality,
 one is obtained from the divergence equation and the second from another Bianchi identity, keeping in mind the identities \eqref{26III23.1}:
\begin{deqarr}
    e_0(\underline \alpha_{ab} ) &=\checkmark&
     \red{
     - e_1( \Phi_{(ab)}) -  2 e_c \underline \Psi^c{}_{(ab)}
     }
      +\lot
    \,,
   \label{23III23.81c}
   \\
   e_0(\underline \beta{}_a) &=
  \checkmark
  &    e_b ( \Phi_a{}^b) +  \lot
  \,,
   \label{23III23.81a}
  \\
  e_1(\underline \beta{}_a) &=\checkmark& - e_b ( \ualpha{}_a{}^b)+  \lot
  \,,
   \label{23III23.81ba}
  \\
  e_0( \beta{}_a) &=\checkmark&   e_b ( \alpha{}_a{}^b)+  \lot
  \,,
   \label{23III23.81bac}
  \\
    e_1( \Phi_{(ab)})
      &=\checkmark&
      \red{
      e_0 (\underline \alpha_{ab})
     +   2 e_{(a} (\underline \beta_{b)})
     }
      +\lot
    \,,
   \label{23III23.81d}
  \\
    e_1 (\underline \Psi_{abc}) &=\checkmark& -e_a ( \underline \alpha_{bc})
     + e_b(\underline \alpha_{ac})
     +\lot
    \,,
   \label{23III23.81h}
  \\
    e_0(\underline \Psi_{abc}) &=\checkmark&  e_b(\Phi_{ca} )  - e_a(\Phi_{cb}) +\lot
    \,,
   \label{23III23.81i}
\end{deqarr}
 Combining \eqref{23III23.81c} and \eqref{23III23.81d} one
finds
 \ptcheck{26III23; but this means that the sigma equation is missing from the system?}
\begin{deqarr}
    e_0(\underline \alpha_{ab} ) &= \checkmark&
     -\red{
        e_{(a} (\underline \beta_{b)})    -   e_c \underline \Psi^c{}_{(ab)}
        }
        +\lot
     \,,
     \label{04IV23.81a}
\\
    e_1(  \Phi_{(ab)} ) &= \checkmark&
    \red{
        e_{(a} (\underline \beta_{b)})
           -   e_c \underline \Psi^c{}_{(ab)}
           }
           +\lot
     \,,
     \label{04IV23.81b}
\end{deqarr}
%
%
%
Similar calculations lead to
\begin{deqarr}
    e_0(\sigma_{ab})
   &=\checkmark&      e_c(  \Psi_{ab}{}^c)
     +\lot
    \,,
   \label{24III23.14}
\\
  e_0(\pi_{abcd})  &=\checkmark&
  e_b (  \Psi_{cda})
  -   e_a(  \Psi_{cdb})
     +\lot
    \,,
   \label{23III23.81k3}
\\
  e_1(\Psi_{abc})
 &=\checkmark&
   - e_0 (\underline \Psi_{abc})
  -  2 e_d(\pi^d{}_{cab})
     +\lot
   \label{23III23.81k4}
\\
     &=\checkmark&
    2 e_c(\sigma_{ab})
   +  e_0 (\underline \Psi_{abc})
     +\lot
    \,,
   \label{23III23.81k5}
\end{deqarr}
Combining the last two equations we obtain
\begin{deqarr}
  e_1(\Psi_{abc})  &=\checkmark&
      e_c(\sigma_{ab})
  -    e_d(\pi^d{}_{cab})
     +\lot
   \label{23III23.71}
\end{deqarr}
%

\section{The null frame decomposition of the Bianchi equations in higher dimensions}
 \label{App28IV23.1}

The aim of this appendix is to present the complete form of the Bianchi equations in higher dimension, as relevant for the arguments in the main body of this paper. We use a null frame as in Section~\ref{sSWcdnf}.
We start with the rotation coefficients.

\subsection{Rotation coefficients}

The {\em null second fundamental forms} of a codimension-two
submanifold $S$
 are the two symmetric tensors on $S$ defined
as
\be \chi(X,Y)=g(D_Xe_0,Y)\,,\qquad \chib(X,Y)=g(D_Xe_1,Y)\,, \eeq
where $D$ is the Levi-Civita connection of $(\M,g)$, while $X,Y$ are
tangent to $S$.  The {\em torsion} of $S$ is a 1-form on $S$,
defined for vector fields $X$ tangent to $S$ by the formula
\be \ze(X)=-\frac{1}{2}g(D_Xe_1,e_0)=\frac{1}{2}g(D_Xe_0,e_1)
 \,.
\eeq
In the definitions above it is also assumed that $e_0$ and $e_1$ are
normal to $S$, so that $\ts$ coincides, over $S$, with the
distribution $TS$ of the planes tangent to $S$.  (Throughout the
indices are raised and lowered with the metric $g$.)

Following%
\footnote{The presentation here follows closely the four-dimensional presentation in~\cite{CCW}, which itself was based on that in~\cite{KlainermanNicoloBook}.}
Klainerman and Nicol\`{o}, we use the following labeling of the
remaining Newman-Penrose coefficients associated with the frame
fields $e_\al$:
\renewcommand{\hxi}{\xi}%
\renewcommand{\hxib}{\underline{\xi}}%
\renewcommand{\heta}{\eta}%
\renewcommand{\hetab}{\underline{\eta}}%
\renewcommand{\home}{\omega}%
\renewcommand{\homb}{\underline{\omega}}%
\renewcommand{\hups}{\upsilon}%
\renewcommand{\hupsb}{\underline{\upsilon}}%
\renewcommand{\dd}{\red{D}}%
\begin{deqarr}
\hxi_{a}&=&\frac{1}{2}g(\dd_{e_0}e_0,e_{a})\,,\label{NP}\\
\hxib_{a}&=&\frac{1}{2}g(\dd_{e_1}e_1,e_{a})\,,\arrlabel{NP.}\\
\heta_{a}&=&-\frac{1}{2}g(\dd_{e_1}e_{a},e_0)
=\frac{1}{2}g(\dd_{e_1}e_0,e_{a})\,,\\
\hetab_{a}&=&-\frac{1}{2}g(\dd_{e_0}e_{a},e_1)
=\frac{1}{2}g(\dd_{e_0}e_1,e_{a})\,,\\
2\home&=&-\frac{1}{2}g(\dd_{e_0}e_1,e_0)
\,, \\
2\homb&=&-\frac{1}{2}g(\dd_{e_1}e_0,e_1)
\,,\\
2\hups&=& - \frac{1}{2}g(\dd_{e_1}e_1,e_0)\,,\\
 2\hupsb&=&- \frac{1}{2}g(\dd_{e_0}e_0,e_1)\,.\label{NPlast}
\end{deqarr}
(The principle that determines which symbols are underlined, and
which are not, should be clear from \Eq{struct1} below: all the
terms \emph{at the right hand side} of that equation have a
coefficient in front of $e_0$ which is underlined.)  The above
definitions, together with the properties of the  connection
coefficients $\hGlocal _{ijk}$, imply the following :
\renewcommand{\hchi}{\chi}%
\renewcommand{\hchib}{\underline{\chi}}%
\renewcommand{\hze}{\zeta}%
\renewcommand{\hzeb}{\underline{\zeta}}%
\renewcommand{\hxi}{\xi}%
\renewcommand{\hxib}{\underline{\xi}}%
\renewcommand{\dd}{\red{D}}%
\begin{deqarr}
\hchi_{ab}& = & \hGlocal _{ab0}= -\hGlocal _{a0b} = 2 \hGlocal
_a{}^1{}_b = - 2 \hGlocal _{ab}{}^1
\,,\nnn\label{gammas0}\\
\hchib_{ab}& = & \hGlocal _{ab1}= -\hGlocal _{a1b} = 2 \hGlocal
_a{}^0{}_b = - 2 \hGlocal _{ab}{}^0
\,,\label{gammasb}\\
\hze_a  & = & \hGlocal _{a}{^1}{}_1 = -\frac 12 \hGlocal _{a01} =
\hGlocal _{a0}{}^0
\,,\nnn\\
\hzeb_a  & = & \hGlocal _{a}{^0}{}_0 = -\frac 12 \hGlocal _{a10} = -
\hGlocal _{a0}{}^0
\,,\nnn\\
\hxi_a  & = &  \hGlocal _0{}^1{}_{a} = - \hGlocal _{0a}{}^1 = \frac
12 \hGlocal _{0a0}= -\frac 12 \hGlocal _{00a}
\,,\nnn\\
\hxib_a  & = &  \hGlocal _1{}^0{}_{a} = - \hGlocal _{1a}{}^0 = \frac
12 \hGlocal _{1a1}= -\frac 12 \hGlocal _{11a}
\,,\nnn\\
\heta_a  & = & \hGlocal _{1}{^1}{}_a = -\frac 12 \hGlocal _{10a} =
\frac 12\hGlocal _{1a0} = -\hGlocal _{1a}{}^1
\,,\nnn\\
\hetab_a  & = & \hGlocal _{0}{^0}{}_a = -\frac 12 \hGlocal _{01a} =
\frac 12\hGlocal _{0a1} = -\hGlocal _{0a}{}^0
\,,\nnn\\
2\home  & = & \hGlocal _{0}{^1}{}_1 = -\frac 12 \hGlocal _{001} =
\hGlocal _{00}{}^0
\,,\nnn\\
2\homb  & = & \hGlocal _{1}{^0}{}_0 = -\frac 12 \hGlocal _{110} =
\hGlocal _{11}{}^1
\,,\nnn\\
2\hups  & = & \hGlocal _{1}{^1}{}_1 = -\frac 12 \hGlocal _{101} =
\hGlocal _{10}{}^0
\,,\nnn\\
2\hupsb  & = & \hGlocal _{0}{^0}{}_0 = -\frac 12 \hGlocal _{010} =
\hGlocal _{01}{}^1 \,. \arrlabel{gammas}
\end{deqarr}
Note that
$$\hGlocal _{\mu}{^0}{}_1 = -\frac 12 \hGlocal _{\mu11}= 0=
\hGlocal _{\mu1}{}^0 \quad \mbox{and}\quad \hGlocal _{\mu}{^1}{}_0 =
-\frac 12 \hGlocal _{\mu00}= 0= \hGlocal _{\mu0}{}^1 $$
The system (\ref{gammas}) leads to
\renewcommand{\hnabb}{\nabb}%
\renewcommand{\hdddd}{\ddd}%
\renewcommand{\hchib}{\chib}%
\renewcommand{\hdivv}{\divv}%
\renewcommand{\hJ}{J}%
\begin{deqarr}
\hDlocal
_{a}e_b&=&\hnabb_{a}e_b+\frac{1}{2}\hchi_{ab}e_1+\frac{1}{2}
        \hchib_{ab}e_0\nnn\,,\\
\hDlocal  _{1}e_a&=&\hdddd_1 e_a+\heta_{a}e_1+\hxib_{a}e_0\nnn\,,\\
\hDlocal  _{0}e_a&=&\hdddd_0 e_a+\hetab_{a}e_0+\hxi_{a}e_1\,,\nnn\\
\hDlocal  _{a}e_1&=&\hchib_{a}{}^{b}e_b+\hze_{a}e_1\nnn\,,\\
\hDlocal  _{a}e_0&=&\hchi_{a}{}^{b}e_b+\hzeb_{a}e_0\nnn\,,\\
\hDlocal  _{1}e_1&=&2\xib^ae_a+2\hups e_1\,,\label{struct1f}\\
\hDlocal  _{0}e_0&=&2 \hxi^ae_a +2\hupsb e_0\,,\nnn\\
\hDlocal  _{0}e_1&=&2\hetab^{b}e_{b}+2\home e_1\,,\nnn\\
\hDlocal  _{1}e_0&=&2\heta^{b}e_{b}+2\homb e_0\,. \arrlabel{struct1}
\end{deqarr}
Here and elsewhere, $\hnabb_{a}e_b$, $\hdddd_1 e_a$ and $\hdddd_0
e_a$ are defined as the orthogonal projection of the left-hand side
of the corresponding equation to $\ts$. We stress that no
simplifying assumptions have been made concerning the nature of the
vector fields $e_a$, except for the orthonormality relations
\eq{metric}.

\subsection{Bianchi equations}

The Bianchi identities, and their contractions in vacuum, provide the following set of equations:

\begin{deqarr}
\label{e0ua}
 \hdddd_0\ua_{ab}&=&\checkmark
-   \hdddd_c\upsi^{c}{}_{(ab)}
 -  \hdddd_{(a}\ub_{b)}
 +      \hze^{c}  \upsi_{c(ab)}
-  4\etab^c\upsi_{c(ab)}
 +4\om\ua_{ab}
-  4{\ub}_{(a}\etab_{b)}
 -  \half    \tr(\chib) \ph_{(ab)}
 \nonumber
 \\
 &&
+      \chib^c{}_{ (a}\ph_{b) c }
- \half \tr(\chi)\ua_{ab}
+    2\chib^c{}_{(a}\si_{b)c}
+     \chib^{cd}\pi_{cabd}
 +    (\hzeb_{(a}+2\hze_{(a} )\ub_{b)}
-  \rho{\chib}_{ab}
 + J_{(ab)1}
  \,,
  \qquad \qquad
\\
%
 \label{e1a}
\hdddd_1\al_{ab}
&=&\checkmark
- \hdddd_c\ps^{c}{}_{(ab)}
+ \hdddd_{(a}\bet_{b)}
 +   \hzeb^{c} \ps_{c(ab)}
- 4\eta^c\ps_{c(ab)}
 + 4\omb\al_{ab}
 +  4\bet_{(a}\eta_{b)}
- \half\tr(\chi)\ph_{(ab)}
\nn
\\
&&
+  \chi^c{}_{(a}\ph_{|c|b)}
- \half\tr(\chib)\al_{ab}
- 2 \chi^c{}_{(a}\si_{b)c}
+    \chi^{cd}\pi_{cabd}
-    (2\hzeb_{(a}+\hze_{(a})\bet_{b)}
-   \rho\chi_{ab}
 + \hJ_{(ab)0}
  \,,
\\
%
 \label{e0ub}
\hdddd_0\ub_a
 &=&\checkmark
 \hdddd_b\ph_{a}{}^b
+(4\om+2\vb -  \tr(\chi))\ub_a +    \chib_{a}{}^b\bet_b
+2 \etab^b \si_{ab}
-   \xi  ^b\ua_{ab}
-  2\rho\etab_a
\nn
\\
&&
- (\hzeb^b+ \hze^b-\etab^b)\ph_{ab}
+  \chib^{bd}\ps_{dab}
 +  \hJ_{01a}
  \,,
\\
  \hdddd_1 \ubeta_a &=&\checkmark
-  \hdddd_{b}\ua^{b}{}_{a}
+(2 \hze^b-\eta^b)\ua_{ba} -    \chib^{b}{}_{a}\ub_{b}
%
\tcp{-    \chib^{bd}\upsi_{dab}}
+(2\omb+4v- \tr(\chib))\ub_a
\nn
\\
&&
 +  ( \ph_{ab} - 2\si_{ba})\xib^b
-  2\rho\xib_a
 -  \hJ_{11a}
  \,,
\label{e1ub}
\\
 \hdddd_0 \beta_a
 &=&  \checkmark \hdddd_b\al_{a}{}^b
 +(\etab^b -2  \hzeb^b)\al_{ab}
-    \chi^b{}_{a}\bet_{b}
%
\tcp{+  \chi^{bd}\ps_{da b}}
 +(2 \om+ 4\vb-\tr(\chi))\bet_a
\nonumber \\
&&
-   \xi ^b(\ph_{ba}
 +  2\si_{ba})
+  2\rho\xi_a
 +  \hJ_{00a}
  \,,
 \label{e0b}
\\
  \hdddd_1 \beta_a
   &=&\checkmark
-  \hdddd_b\ph^{b}{}_{a}
+ ( \hze^b+ \hzeb^b -
 \eta^b )\ph_{ba} +
 2\eta^b \si_{ab}
+(4\omb
 +   2v - \tr(\chib))\bet_a
\nonumber
\\
&&
-    \chi ^{bd}\upsi_{dab}
 +    \chi^{b}{}_{a}\ub_b
+   \xib^b\al_{ba}
+ 2\eta_a\rho
\nonumber
\\
&&
 -  \hJ_{10a}
 \label{e1b}
\\
\label{e0p}
 \hdddd_0  \ph_{ab}
&=&\checkmark
-  \hdddd_a   \bet_{b}
-  \hdddd_c\ps^{c}{}_{ab}
+ 2 (\om+\vb)\ph_{ab}
+    2\xi_b\ub_a
+    2\xi^c \upsi_{bca}
 -   \half \tr(\chi)\ph_{ab}
\nonumber
\\
&&
- \half\tr( \chib)\al_{ab}
+ 2   \chi^c{}_{[a}\si_{b]c}
+    \tcp{ \chib^c{}_{a}}\al_{bc} \nn
 +  (\hzeb^{c}-2\etab^c)\ps_{cab}
\nonumber
\\
&&
+     \chi^{cd}\pi_{cabd}
+   \rho\chi_{ab}
+   (2\hzeb_a+\hze_a-2\etab_a)\bet_{b}
+ \hJ_{ab0}
 \\
   \hdddd_1 \ph_{ab} &=&\checkmark -   \hdddd_c\upsi^{c}{}_{ba}
+  \hdddd_b    \ub_{a}
 +     (\hze^{c}-  2\eta^c) \upsi_{cba}
+  2(\omb+v)\ph_{ab}
 -    2\xib_a\bet_b
+  2 \xib^c \ps_{acb}
 \nonumber
\\
&&
 -   \half  \tr( \chib)\ph_{ab}
- \half \tr( \chi)\ua_{ab}
 +  \chi^c{}_{ b}\ua_{c a}
 -    \chib^c{}_{a}\si_{cb}
+     \chib^{cd}\pi_{cbad}
 \nonumber
\\
&&
  -    (\hzeb_b+2\hze_b -2\eta_b)\ub_a
+  \rho{\chib}_{ba}
-   \chib_{b}{}^c\si_{ac}
 + J_{ba1}
  \,,
 \label{e1p}
\\
  \hdddd_0  \pi_{abcd}
&=&\checkmark
- \hdddd_a   \ps_{cdb}
+\hdddd_b    \ps_{cda}
-  2\etab_{[a}\ps_{|cd|b]}
- 2 \xi_{[a}\upsi_{|cd|b]}
-  2\etab_{[c}\ps_{|ab|d]}
\nn
\\
&&
- 2 \xi_{[c}\upsi_{|ab|d]}
+   2 \hzeb_{[a} \ps_{|cd|b]}
+  \chi_a{}^e \pi_{bedc}
-    \chib_{a[c}\al_{d]b}
-    \chi_{a[c}\ph_{d]b}
\nn
\\
&&
 +  \chi _{b}{}{^e}\pi_{eacd}
+  \chib_{b[c} \al_{d]a}
+  \chi_{b[c} \ph_{d]a}
  \,,
\label{e0pi}
\\
\tcp{\hdddd_1 \pi_{abcd}}
&=&
- \hdddd_a   \upsi_{cdb}
+\hdddd_b    \upsi_{cda}
-  2\eta_{[a}\upsi_{|cd|b]}
- 2 \xib_{[a}\ps_{|cd|b]}
-  2\eta_{[c}\upsi_{|ab|d]}
\nn
\\
&&
- 2 \xib_{[c}\ps_{|ab|d]}
+   2 \hze_{[a} \upsi_{|cd|b]}
+  \chib_a{}^e \pi_{bedc}
-    \chi_{a[c}\ua_{d]b}
-    \chib_{a[c}\ph_{|b|d]}
\nn
\\
&&
+  \chib _{b}{}{^e}\pi_{eacd}
+  \chi_{b[c} \ua_{d]a}
+  \chib_{b[c} \ph_{|a|d]}
\,,
\label{e1pi}
\\
 \hdddd_0  \upsi_{abc} &=&\checkmark
- \hdddd_a   \ph_{cb}
 + \hdddd_b    \ph_{ca}
- 2\etab_{[a}\ph_{|c|b]}
 -  2\xi_{[a}\ua_{b]c}
-  2\etab_c\si_{ab}
\nn
\\
&&
+  2\om \upsi_{abc}
+ 2\etab^d\pi_{abcd}
+  2\hzeb_{[a}\ph_{|c|b]}
-  2\chib_{[a|c|}\bet_{b]}
+  2\hze_{[a}\ph_{|c|b]}
\nn
\\
&&
 + \chi_{b}{}^d\upsi_{dac}
+   \chi_{a}{}^d \upsi_{bdc}
- \chib_b{}^d\ps_{cda}
+ \chib_a{}^d\ps_{cdb}
  \,,
 \label{e0ups}
\\
\hdddd_0  \ps_{abc}
&=&\checkmark
-\hdddd_a   \al_{bc}
+ \hdddd_b    \al_{ac}
- 2 \etab_{[a}\al_{b]c}
- 2\xi_{[a}\ph_{b]c}
+  2\xi_c\si_{ab}
+    2\vb \ps_{abc}
+ 2\xi^d\pi_{abcd}
 \nn
 \\
 &&
+  4\hzeb_{[a}\al_{b]c}
-  2 \chi_{a}{}^d \ps_{d(bc)}
+ 2\chi_{b}{}^d\ps_{d(ac)}
-  2\chi_{b(a}  \bet_{c)}
 + 2\chi_{a(b}\bet_{c)}
  \,,
\label{e0ps}
\\
\label{e1ups}
 \hdddd_1 \upsi_{abc}
&=&\checkmark
-\hdddd_a   \ua_{bc}
+\hdddd_b    \ua_{ac}
-  2\xib_{[a}\ph_{|c|b]}
-  2\eta_{[a}\ua_{b]c}
\nn
-  2\xib_c\si_{ab}
+  2v\upsi_{abc}
+ 2\xib^d\pi_{abcd}
\\
&&
 +   4\hze_{[a}\ua_{b]c}
 - 2\chib_{a(b}\ub_{c)}
+ 2 \chib_{b(a}\ub_{c)}
+ 2 \chib_{b}{}^d\upsi_{d(ac)}
-2 \chib _{a}{}^d \upsi_{d(bc)}
  \,,
\\
 \hdddd_1 \ps_{abc}
&=&\checkmark
- \hdddd_a   \ph_{bc}
+ \hdddd_b    \ph_{ac}
 -  2\xib_{[a}\al_{b]c}
- 2 \eta_{[a}\ph_{b]c}
+  2\eta_c\si_{ab}
+ 2\omb\ps_{abc}
+ 2\eta^d\pi_{abcd}
\nn
\\
&&
 +  2\hze_{[a}\ph_{b]c}
+ 2\chi_{[a|c|}\ub_{b]}
+2 \hzeb_{[a}\ph_{b]c}
\nn
\\
&&
+ \chib_{b}{}^d\ps_{dac}
-  \chib_{a}{}^d \ps_{dbc}
- \chi_b{}^d\upsi_{cda}
+ \chi_a{}^d\upsi_{cdb}
\end{deqarr}
and
\begin{deqarr}
 \label{e1psb}
 \hdddd_1 \ps_{abc}
&=&\checkmark
  \hdddd_c    \si_{ab}
-   \hdddd_d    \pi^{d}{}_{cab}
 + 2 \omb\ps_{abc}
\tcp{+  2 \eta^d \pi_{abcd} }
 +   (2\eta_c-\hzeb_c-\hze_c)\si_{ab}
\nn
\\
&&
 -   2 \xib_{[a}\al_{b]c}
 - 2 \eta_{[a}\ph_{b]c}
-      \chib^{d}{}_{[d}\ps_{|ab|c]}
-    \chib^{d}{}_{[a}\ps_{|dc|b]}
-    \chi^d{}_{[d}\upsi_{|ab|c] }
 -   \chi^d{}_{[a}\upsi_{|dc|b]}
\nn
\\
&&
+  \chib_{c[a}\bet_{b]}
 +   \chi_{c[a}\ub_{b]}
 +  \half\chi_c{}^d\upsi_{abd}
- \half\chib_c{}^d  \ps_{abd}
 + \hJ_{cab}
  \,,
\\
 \label{e0upsb}
 \hdddd_0  \upsi_{abc}
&=&\checkmark
-  \hdddd_c    \si_{ab}
-   \hdddd_d    \pi^{d}{}_{cab}
+   2\om\upsi_{abc}
\tcp{+  2 \etab^d \pi_{abcd} }
+   (\hzeb_c+\hze_c-2\etab_c)\si_{ab}
\nn
\\
&&
 -  2\xi_{[a}\ua_{b]c}
-   2 \etab_{[a}\ph_{|c|b]}
-      \chib^{d}{}_{[d}\ps_{|ab|c]}
-    \chib^{d}{}_{[a}\ps_{|dc|b]}
-    \chi^d{}_{[d}\upsi_{|ab|c] }
-   \chi^d{}_{[a}\upsi_{|dc|b]}
\nn
\\
&&
-   \chib_{c[a}\bet_{b]}
 -   \chi_{c[a}\ub_{b]}
-  \half\chi_c{}^d\upsi_{abd}
+\half \chib_c{}^d  \ps_{abd}
 + \hJ_{cab}
\\
\label{e0r} e_0(\rho) &=& \checkmark  \hdivv \beta-
 \rho\tr(\chi) +4\rho(\vb+\omega)-2\hxi\cdot \ub
+(2 \hetab -2\hzeb -\hze )\cdot \beta \nonumber
\\
&&
  -\half\al_{bc}\uchi^{bc}
 +\half\ph_{cb}\chi^{bc}
  +\half \hJ_{001}
\\
e_1(\rho)&=&\checkmark -\hdivv \ub - \tr \hchib\,\ro
 +2 \hxib\cdot \beta
+4 (\homb+ \hups)\rho
 +   \left(-2\eta +\hzeb +2\ze \right)\cdot
\ub
 \nonumber
 \\
 &&
-\frac 12 h^{ab}h^{ce}\chi_{ae}\ua_{bc}+\half
h^{ab}h^{ce}\chib_{ae}\ph_{bc}  +\frac 12 \hJ_{110}
\label{e1r}
\\
 \label{e0s}
 \hdddd_0  \si_{ce} &=&\checkmark
   \hdddd_b\ps_{ce}{}^b
+(\etab^b-   \hzeb^b ) \ps_{ceb}
%
%
+    \chib^{b}{}_{[c}\al_{|b|e]}
+  \chi^b{}_{[c}\ph_{e]b}
\nonumber
\\
&&
 +( 2 \om+
 2\vb -\tr(\chi))\si_{ce}
-  \xi ^b\upsi_{ceb}
-  2\etab_{[c}\bet_{e]}
-   2\xi_{[c}\ub_{e]}
 -  \hJ_{0ce}
  \,.
\nn
\\
\label{e1s}
 \hdddd_1 \si_{ce}
 &=&\checkmark
-   \hdddd_b\upsi_{ce}{}^b +   (\hze^b-\eta^b)\upsi_{ceb}
+ \underbrace{ \chib^{bd}\pi_{bdce}}_{=0}
  - \chib^b{}_{[c}\ph_{|b|e]}
 -   \chi^b{}_{[c}\ua_{|b|e]}
 \nonumber
 \\
 &&
 +   (2 \omb+ 2v - \tr(\chib))\si_{ce}
+   \xib ^b\ps_{ceb}
  -  2\xib_{[c}\bet_{e]}
-  2\eta_{[c}\ub_{e]}
\nonumber
\\
&& +  \hJ_{1ce}\,,
\\
\label{e0s2}
\tcp{\hdddd_0  \si_{ab}} &=&2 \big(-\hdddd_{[a} \bet_{b]} +(\etab+ \zetab)_{[a}\bet_{b]} -\xi_{[a}\ub_{b]} +(\eta+\xi)^c\Psi_{c[ab]}  \nn\\
&&
-\half\chib_{[a}{}^c\al_{b]c} -\half \chi_{[a}{}^c\Phi_{|c|b]} + \chi_{[a}{}^c\sigma_{b]c} \big)\,,\\
\label{e1s2}
\tcp{\hdddd_1 \si_{ab}} &=&2 \big(-\hdddd_{[a} \ub_{b]} +(\eta+ \zeta)_{[a}\ub_{b]} -\xib_{[a}\bet_{b]} +(\etab+\xib)^c\Psib_{c[ab]}  \nn\\
&&
-\half\chi_{[a}{}^c\ab_{b]c} + \half \chib_{[a}{}^c\Phi_{b]c} + \chib_{[a}{}^c\sigma_{b]c} \big)\,.
\label{Syst}
 \end{deqarr}
Note that the right-hand sides of the  equations for $\al$, \eqref{e1a},  and $\ua$,  \eqref{e0ua}, are traceless when taking account of the relations \eqref{26III23.1}.

All equations have been checked with the open source computer algebra software Cadabra~\cite{Peeters:2006kp,Peeters:2007wn,Peeters:2018dyg}.

\bibliographystyle{amsplain}
\bibliography{ChruscielWafoFinn-minimal}

\providecommand{\bysame}{\leavevmode\hbox to3em{\hrulefill}\thinspace}
\providecommand{\MR}{\relax\ifhmode\unskip\space\fi MR }
\providecommand{\MRhref}[2]{%
  \href{http://www.ams.org/mathscinet-getitem?mr=#1}{#2}
}
\providecommand{\href}[2]{#2}
\begin{thebibliography}{10}

\bibitem{CCW}
A.~Cabet, P.T. Chru\'{s}ciel, and R.~Tagne~Wafo, \emph{On the characteristic
  initial value problem for nonlinear symmetric hyperbolic systems, including
  {E}instein equations}, Dissertationes Math.\ (Rozprawy Mat.) \textbf{515}
  (2016), 72 pp., arXiv:1406.3009 [gr-qc]. \MR{3528223}

\bibitem{CCM2}
Y.~Choquet-Bruhat, P.T. Chru\'{s}ciel, and J.M. Mart\'in-Garc\'ia, \emph{{The
  Cauchy problem on a characteristic cone for the Einstein equations in
  arbitrary dimensions}}, Ann.\ H.\ Poincar\'e \textbf{12} (2011), 419--482,
  arXiv:1006.4467 [gr-qc]. \MR{2785136}

\bibitem{ChristodoulouKlainerman93}
D.~Christodoulou and S.~Klainerman, \emph{The global nonlinear stability of the
  {M}inkowski space}, Princeton Mathematical Series, vol.~41, Princeton
  University Press, Princeton, NJ, 1993. \MR{MR1316662 (95k:83006)}

\bibitem{Collingbourne}
S.~Collingbourne, \emph{The {Gregory–Laflamme} instability and conservation
  laws for linearised gravity}, Ph.D. thesis, University of Cambridge, 2022,
  \url{
  https://api.repository.cam.ac.uk/server/api/core/bitstreams/6a5215f2-5719-4ca4-8f28-7a4131482097/content}.

\bibitem{ReallHighDimNP}
M.~Durkee, V.~Pravda, A.~Pravdova, and H.S. Reall, \emph{{Generalization of the
  Geroch-Held-Penrose formalism to higher dimensions}}, Class.\ Quantum Grav.
  \textbf{27} (2010), 215010, arXiv:1002.4826 [gr-qc].

\bibitem{FriedrichCargese}
H.~Friedrich, \emph{Smoothness at null infinity and the structure of initial
  data}, The Einstein Equations and the Large Scale Behavior of Gravitational
  Fields (P.T. Chru\'{s}ciel and H.~Friedrich, eds.), Birkh{\"a}user, Basel,
  2004, pp.~121--203, arXiv:gr--qc/0304003.

\bibitem{KlainermanNicoloBook}
S.~Klainerman and F.~Nicol{\`o}, \emph{The evolution problem in general
  relativity}, Progress in Mathematical Physics, vol.~25, Birkh{\"a}user,
  Boston, MA, 2003. \MR{1 946 854}

\bibitem{Luk}
J.~Luk, \emph{On the local existence for the characteristic initial value
  problem in general relativity}, Int.\ Math.\ Res.\ Not.\ IMRN (2012),
  4625--4678. \MR{2989616}

\bibitem{PravdaHighDimNP}
M.~Ortaggio, V.~Pravda, and A.~Pravdova, \emph{{Algebraic classification of
  higher dimensional spacetimes based on null alignment}}, Class. Quant. Grav.
  \textbf{30} (2013), 013001, arXiv:1211.7289 [gr-qc].

\bibitem{Peeters:2006kp}
K.~Peeters, \emph{{A field-theory motivated approach to symbolic computer
  algebra}}, Comput.\ Phys.\ Commun. \textbf{176} (2007), 550--558,
  arXiv:cs/0608005.

\bibitem{Peeters:2007wn}
\bysame, \emph{{Introducing Cadabra: A Symbolic computer algebra system for
  field theory problems}},  (2007), arXiv:hep-th/0701238.

\bibitem{Peeters:2018dyg}
\bysame, \emph{{Cadabra2: computer algebra for field theory revisited}}, Jour.\
  Open Source Softw. \textbf{3} (2018), no.~32, 1118.

\bibitem{RendallCIVP2}
A.D. Rendall, \emph{The characteristic initial value problem for the {E}instein
  equations}, Nonlinear hyperbolic equations and field theory ({L}ake {C}omo,
  1990), Pitman Res. Notes Math.\ Ser., vol. 253, Longman Sci. Tech., Harlow,
  1992, pp.~154--163. \MR{MR1175208 (93j:83010)}

\bibitem{RodnianskiShlapentokh}
I.~Rodnianski and Y.~Shlapentokh-Rothman, \emph{The asymptotically self-similar
  regime for the {E}instein vacuum equations}, Geom.\ Funct.\ Anal. \textbf{28}
  (2018), 755--878. \MR{3816523}

\bibitem{SenovillaSymHyp}
J.M.M. Senovilla, \emph{Symmetric hyperbolic systems for a large class of
  fields in arbitrary dimension}, Gen.\ Rel.\ Grav. \textbf{39} (2007),
  361--386. \MR{2322655}

\end{thebibliography}

\end{document}